# Lossless integrated RF photonic filter with record-low noise figure and 116 dB of dynamic range

Yang Liu[1,2], Jason Hotten[1,2], Amol Choudhary[1,2], Benjamin J. Eggleton[1,2], and David Marpaung[1,2*]

[1]Centre for Ultrahigh bandwidth Devices for Optical Systems (CUDOS), University of Sydney, Sydney, Australia

[2]Australian Institute for Nanoscale Science and Technology (AINST), University of Sydney, Sydney, Australia

*E-mail: david.marpaung@sydney.edu.au

*Abstract*— We report a silicon nitride RF photonic notch filter with unprecedented performance including 8 dB of RF gain, a record-low noise figure of 15.6 dB, and a spurious-free dynamic range of 116 dB while achieving an ultra-deep rejection of beyond 50 dB in the stopbands.

## I. Introduction

The incorporation of photonic integrated circuits in microwave photonic systems offers tremendous potential including massive reduction of footprint as well as enhanced functionalities and stability [1]. Recent advances in this field have the miniaturization and integration of a variety of advanced RF photonic signal processing subsystems such as tunable filters [1,2]. But to find practical applications in RF and microwave systems, these integrated microwave photonic devices need to satisfy a number of stringent performance metrics including low insertion loss, high signal-to-noise ratio, and high linearity [3]. In RF photonics, these requirements translate to high RF link gain, low noise figure (NF), and high spurious-free dynamic range (SFDR). Thus far, achieving on-chip functionalities with these stringent requirements have been proven challenging and advances in this area are limited. In virtually all reported cases, the performance of integrated MWP systems are prohibitively low, typically with 20 dB of RF loss [2], 40 dB of NF [3], and SFDR of 81 dB.Hz$^{2/3}$ [2]. This is in stark contrast with what is achievable in state-of-the-art MWP links *without* functional photonic circuits that exhibit RF gain of >10 dB, NF of 6-10 dB, and SFDR of >120 dB.Hz$^{2/3}$ [4]. The insertion of a functional photonic circuit leads to excessive losses and incompatibility with techniques commonly used to enhance link performance, for example low biasing [3]. This massive gap between the performance of integrated MWP systems and MWP links is currently left unaddressed.

In this work, we report the first integrated MWP filter with optimized RF performance, exhibiting an RF gain of 8 dB, a record-low NF of 15.6 dB, and a high SFDR of 116 dB.Hz$^{2/3}$. This unparalleled performance is simultaneously obtained with advanced filtering functions featuring dual independently-tunable stop-bands, each with > 50 dB rejection and high spectral resolution of 500 MHz. To further signify the filter performance we emulate RF spectrum filtering of a signal in the presence of an interferer. We show that the filter amplifies the desired signal by 2 dB while attenuating the interferer by 47 dB. Key to these breakthrough results is the use of a low-loss silicon nitride photonic circuit consisting of series of a ring resonators as pre-processing and optical filtering units [5], along with low biasing of a Mach-Zehnder modulator as the link optimization technique to achieve low NF. These results reveal, for the first time, ways of achieving all-optimized integrated RF photonic filter without any performance trade-off and point to the feasibility of implementing integrated MWP sub-systems in RF front ends for real-world applications.

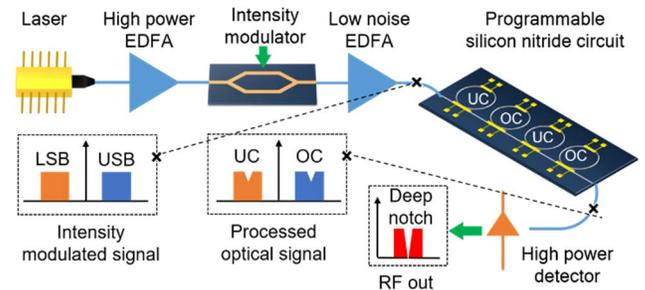

Fig. 1 Schematic of the integrated RF photonic filter. LSB: lower sideband, USB: upper sideband, UC: under-coupled, OC: over-coupled.

## II. Experiments and Results

The schematic of the integrated MWP filter is shown in Fig. 1. An optical carrier from a laser (Teraxion 16.4 dBm ouput power) is amplified using an erbium-doped optical amplifier (EDFA, Amonics) to 30 dBm optical power. The carrier was then intensity-modulated using a Mach-Zehnder modulator (EOSpace 20 GHz MZM) with 3.8 V half-wave voltage and 4 dB of insertion loss. As discussed later, the bias voltage of the MZM was optimized to achieve the lowest RF noise figure. The output of the MZM is an optical spectrum with equal-amplitude and in-phase RF sidebands. This signal is then passed through a low-noise EDFA with 21 dBm of output power before being injected to a programmable silicon nitride circuit (Satrax BV) consisting of four optical ring resonators connected in series. The chip was equipped with on-chip tapers to minimize fiber-to-fiber insertion loss to 7.5 dB. The propagation loss of the optical waveguides is less than 0.2 dB/cm. The rings have a free spectral range (FSR) of 25 GHz and their coupling coefficients and resonance frequencies are tunable through thermo-optic tuning.

To form an RF bandstop filter with a single stopband, two ring resonators are used in series. One ring is used as a pre-processing unit operating in an over-coupling regime and is aligned with the frequencies of the upper sideband. The second ring is operated in the under-coupling regime and is used to filter the upper sideband (Fig. 1). The two rings are tuned to exhibit the same peak attenuation, but because of the different coupling conditions, they exhibit entirely different phase response and quality factor. The over-coupled (OC) ring produces a phase shift of $\pi$ at the center frequency due to its inherent phase transition from 0 to $2\pi$, while the under-coupled (UC) ring exhibits no phase-jump [5]. This precise amplitude and phase tailoring allows for complete destructive interference at the notch frequency in the RF spectrum upon photodetection (Finisar HPDV2120R). Equally important, outside the notch frequencies the mixing products of the RF sidebands will add constructively, hence forming a strong passband. A second independently-tunable stopband was subsequently added by activating the third and fourth rings with similar settings.

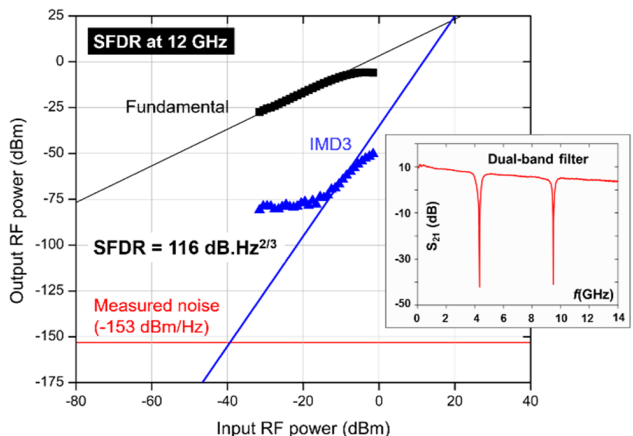

Fig. 2. Measured dynamic range at 12 GHz. Inset: filter transfer function with dual independently tunable passbands.

Unlikel previously reported filters, stopband formation in this technique is truly independent of the MZM bias. This means that the MZM bias can be used separately to optimize the noise figure and link gain of the filter. By extremely low-biasing the MZM at the bias angle of $0.05\pi$, we observe an RF gain of ~8 dB and a minimized NF of 15.6 dB at the frequency of 2 GHz.

We performed a two-tone test with 10 MHz tone spacing and measured the 2$^{nd}$ and 3$^{rd}$-order intercept points and dynamic range at various RF frequencies. The measured 3$^{rd}$-order input intercept point (IIP3) at the frequency of 12 GHz was 19 dBm, resulting in an SFDR of 116 dB.Hz$^{2/3}$. Fig. 2 depicts the measured SFDR at 12 GHz, with and inset showing the dual-stopband filter each with 50 dB of suppression and 500 MHz spectral resolution. The measured RF gain, NF, and SFDR over the entire tuning range of the filter (1-12 GHz) are shown in Fig. 3. The filter consistently achieved RF amplification over the entire band, with NF variations of 15.6 -19.5 dB and SFDR variation of 113-116 dB.Hz$^{2/3}$.

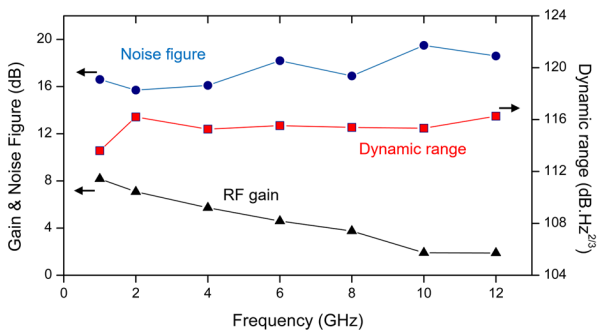

Fig.3. The measured RF gain, noise figure and dynamic range over the entire tuning range of the filter.

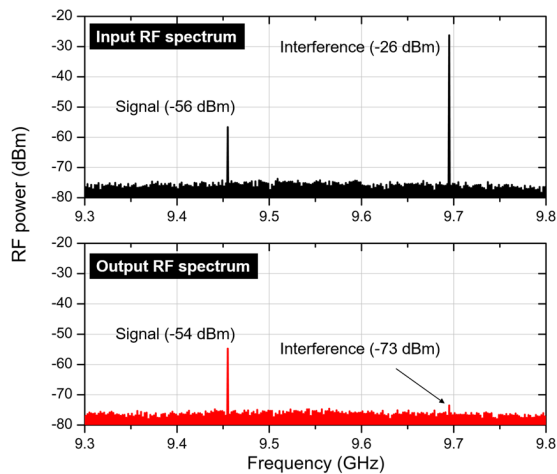

Fig. 4 High resolution RF filtering using the integrated RF photonic filter.

Finally, we emulate transmission of a relatively weak RF signal in the presence of a strong interferer using the filter. The result was shown in Fig. 4. After passing through the filter, the desired signal (9.46 GHz) was amplified by 2 dB, while the interferer which is located 240 MHz away from the signal was reduced by 47 dB. This result highlights the unique performance of the filter in terms of passband amplification, stopband rejection, and spectral resolution.

## III. CONCLUSIONS

We have demonstrated the first integrated MWP filter with simultaneously all-optimized performance. The filter exhibits record-low noise figure, high dynamic range, amplification in the passband and ultrahigh stopband rejection. Our results point to the first step towards widespread applications of integrated MWP systems.

ACKNOWLEDGMENT

Australian Research Council for grants DE150101535, DE170100585, FL120100029, AOARD FA2386-16-1-4036.